\documentclass[12pt]{article}
\usepackage{amssymb,graphicx}
\usepackage{epstopdf}
\usepackage{amsmath,amsfonts}
\usepackage{epsfig}

\def\d{\partial}

\newcommand{\be}{\begin{equation}}
\newcommand{\ee}{\end{equation}}
\newcommand{\bea}{\begin{eqnarray}}
\newcommand{\eea}{\end{eqnarray}}
\newcommand{\bg}{\begin{gather}}
\newcommand{\eg}{\end{gather}}
\newcommand{\bseq}{\begin{subequations}}
\newcommand{\eseq}{\end{subequations}}

\renewcommand{\ln}{\mathop{\rm ln}\nolimits}

\begin{document}
\begin{flushright}
\end{flushright}
\vspace{10pt}
\begin{center}
  {\LARGE \bf Phantom dark energy\\ [0.3cm] with tachyonic instability:\\
[0.3cm] metric perturbations } \\
\vspace{20pt}
A. Sergienko$^{a,b}$, V. Rubakov$^{b}$\\
\vspace{15pt}
  $^a$\textit{
Physics Department, Moscow State University
  }\\
\vspace{5pt}
$^b$\textit{
Institute for Nuclear Research of
         the Russian Academy of Sciences,\\  60th October Anniversary
  Prospect, 7a, 117312 Moscow, Russia}

    \end{center}
    \vspace{5pt}

\begin{abstract}
We study the behavior of metric perturbations in a recently
proposed model of phantom dark energy with tachyonic instability
at long wavelengths. We find that
metric perturbations exponentially grow in time, starting
from very small values determined by vacuum fluctuations,
and may become sizeable at late times. This property may be
of interest for phenomenology.
\end{abstract}

\section{Introduction}

Among models of dark energy (for reviews see, e.g.,
Refs.~\cite{Sahni:1999gb,Carroll:2000fy,Padmanabhan:2002ji,Peebles:2002gy,Sahni:2006pa,Nojiri:2006ri,Copeland:2006wr}),
perhaps the most exotic ones are those exhibiting phantom equation of state,
$p=w\rho$ with $w<-1$. Observational data strongly constrain this
possibility~\cite{Astier:2005qq,Riess:2006fw,WoodVasey:2007jb,Cole:2005sx,Tegmark:2006az,Komatsu:2008hk},
but by no means rule it out, especially if the equation of state parameter
$w$ varies in time (which is the case in many phantom models).
For phantom matter, weak energy condition is violated, signaling
instabilities and possibly other pathologies. Indeed, the simplest phantom
models like those involving negative energy scalar field~\cite{Caldwell:1999ew}
have ghosts in the spectrum at arbitrarily high spatial momenta, and hence are
plagued by catastrophic vacuum instability unless one introduces Lorentz-violating
cutoff into the theory~\cite{Carroll:2003st,Cline:2003gs}.

This situation is not generic, however. First, phantom behavior may be mimicked
in scalar-tensor and $f(R)$
gravities~\cite{Boisseau:2000pr,Gannouji:2006jm,Amendola:2006we}
which need not have instabilities at all. Second, in theories with
Lorentz-violating phantom fields, the violation of weak energy condition
at cosmological distances does not imply dangerous instabilities
at much shorter length scales. Indeed, phenomenologically
acceptable models have been
proposed~\cite{Senatore:2004rj,Creminelli:2006xe,Rubakov:2006pn,Libanov:2007mq}
that have precisely this  property. In particular, the model of
Refs.~\cite{Rubakov:2006pn,Libanov:2007mq} has tachyon in the spectrum of
small perturbations about homogeneous Lorentz-violating phantom background,
but the instabilities occur
at low spatial momenta only, so that their time scale
may be roughly comparable to (though somewhat smaller than)
the present age of the Universe.

The latter property may well be generic to a subclass of phantom
models, in which phantom instability is of tachyonic nature. So,
it is of interest to understand, using the model of
Refs.~\cite{Rubakov:2006pn,Libanov:2007mq} as a prototype, what
observable consequences the long-wavelength tachyonic
instabilities may have. As a step in this direction, we calculate
in this paper the  metric perturbations
generated by the tachyon modes in this model. We find no surprise:
like the tachyon modes themselves, metric perturbations
exponentially grow in time, starting from very small values
determined by the quantum physics of vacuum fluctuations. This may
give rise to potentially observable effects, as the Universe
rapidly becomes less homogeneous at late times.

\section{The model}

The model for dark energy
we study in this paper is the Einstein gravity interacting with
a vector field $B_{\mu}$ and a scalar field $\phi$.
Its construction combines two ideas. One~\cite{Gripaios:2004ms} is that
in the absence of gauge invariance,
a theory of a vector field may nevertheless have healthy spectrum,
provided that the background vector field is non-trivial. The
other~\cite{Libanov:2005vu} has to do with the fact that with vector field(s),
one can construct generally covariant one-derivative terms in the Lagrangian,
that dominate at low momenta and become negligible at high momenta. Hence,
the
Lagrangian for the vector and scalar fields is~\cite{Rubakov:2006pn}
\begin{equation}
\emph{L} = -\frac{1}{2} \alpha(\xi) D_\mu B_\nu D^\mu B^\nu +
\frac{1}{2} \beta(\xi) D_\mu B_\nu D^\mu B_\lambda \frac{B^\nu
B^\lambda}{\Lambda^2} + \frac{1}{2} \partial_\mu \phi
\partial^\mu \phi + \epsilon \partial_\mu \phi B^\mu + \frac{M^2}{2} B_\mu B^\mu - \frac{m^2}{2}
\phi^2 \; ,
\nonumber
\end{equation}
where $\xi=B_\mu B^\mu/\Lambda^2$, $\Lambda$ is an UV cut-off scale,
dimensionless parameters $\alpha$ and $\beta$ are
functions of $\xi$, and $\epsilon$ is a positive constant
of dimension of mass.
Upon introducing the
notation $B^2=B_\mu B^\mu$, this Lagrangian
can be rewritten as follows,
\begin{equation} \label{lagrangian}
\emph{L} = -\frac{\alpha}{2} D_\mu B_\nu D^\mu B^\nu +
\frac{\alpha+\gamma}{8} \frac{\partial_\mu
\left(B^2\right)\cdot\partial^\mu \left(B^2\right)}{B^2} +
\frac{1}{2}\partial_\mu\phi\partial^\mu\phi +
\epsilon\partial_\mu\phi B^\mu + \frac{M^2}{2} B^2 - \frac{m^2}{2}
\phi^2 \; ,
\end{equation}
where $\gamma(\xi)=\frac{B^2}{\Lambda^2}
\beta(\xi)-\alpha(\xi)$. To simplify formulas,
in what follows we  assume that
$\alpha(\xi)=\mbox{const}$, $\gamma(\xi)= \mbox{const}$.
The range of parameters in which the model exhibits phantom behavior
is
\begin{equation} \label{approx}
M \lesssim m \ll \epsilon \; ,
\end{equation}
and $\alpha \sim \gamma$.
Spatially flat homogeneous background in this model is described by
\[
g_{\mu\nu}=a^2(\eta) \eta_{\mu\nu}\; ,
\;\;\;\; B_0=a(\eta) X (\eta)\; , \;\;\;\; \phi =\varphi (\eta) \; ,
\]
where $\eta$ is conformal time. Importantly, temporal component of the
vector field does not vanish, which ensures that the expression
(\ref{lagrangian}) makes sense, and that the spectrum of perturbations
about this background is healthy~\cite{Rubakov:2006pn,Libanov:2007mq}
at high spatial momenta
(no ghosts, tachyons or superluminally
propagating modes),
provided that
\[
  \alpha > \gamma > 0 \; .
\]
As we will see shortly, one of the scalar modes of perturbations becomes
a tachyon at sufficiently low momenta.

Cosmology in this model,
with  the usual and dark
matter added, is fairly interesting~\cite{Libanov:2007mq}.
In a wide range of initial data for $X$ and $\varphi$,
these fields stay constant
at early, matter dominated stage, and the corresponding equation of
state is $w=-1$. Then they start evolving, the dark energy
equation of state first being normal, $w>-1$, and later phantom,
$w<-1$. The evolution often occurs in the slow roll regime, so that
$|w+1|$ is  small in the entire course of the cosmological expansion.
Finally, the
system approaches the de~Sitter attractor, which to the leading order in
$m/\epsilon$, $M/\epsilon$ is characterized by the following values
of the Hubble
parameter and fields,
\be
H_A =\frac{M}{\sqrt{3\alpha}} \; , \;\;\;\;
 X_A =-\frac{m M_{PL}}{\sqrt{12\pi}\epsilon} \; ,\; \;\;\;
 \varphi_A=\frac{M M_{PL}}{\sqrt{4\pi\alpha}m} \; .
\nonumber
\ee
The interesting regime in which dark energy
contributes substantially
to the total energy density, occurs at
\be
    H \sim H_A \; , \;\;\;\; X \sim X_A \; , \;\;\; \varphi \sim \varphi_A \; .
\label{mar15-1}
\ee
In what follows, we will be interested precisely in this range of
the cosmological variables.

\section{Tachyonic perturbations}

The linearized perturbations of dark energy fields break up into
vector and scalar parts. The vector sector is healthy at all
spatial momenta, whereas the scalar sector has tachyons at
relatively low momenta.
Hence, we concentrate on the scalar sector.
In Minkowski space-time, the momenta at which one of the modes is tachyonic
are of order $P \sim \epsilon$,
and the corresponding ``frequencies'' are of the same
order~\cite{Rubakov:2006pn,Libanov:2007mq}. In the
expanding Universe, a mode of a given conformal momentum $p$
becomes tachyonic as the physical momentum redshifts down to
$P\equiv p/a \sim \epsilon$. Our purpose is to calculate the
metric perturbations generated by the tachyonic modes\footnote{In fact,
the behavior of modes in Minkowski space-time at $P \lesssim M$
is somewhat more complicated~\cite{Libanov:2007mq}. This region
of momenta is not of interest for our purposes, as it corresponds
to physical momenta, and hence frequencies, small compared to the Hubble
parameter, see (\ref{mar15-1}).}.

With scalar perturbations included, the fields and metric in
conformal Newtonian gauge are given by \bea B_0
(\eta,\textbf{x})=a(\eta)X(\eta)&+&b_0(\eta,\textbf{x})\;
,\;\;\;\; B_i(\eta,\textbf{x})=b_i(\eta,\textbf{x})\; ,\;\;\;\;
\phi(\eta,\textbf{x})= \varphi(\eta) +
\frac{\chi(\eta,\textbf{x})}{a(\eta)}
\nonumber \\
ds^2 &=& a^2(\eta)\left[(1 + 2\Phi) d\eta^2 - (1-2\Phi) d{\bf x}^2
\right] \nonumber \eea where $b_i$ is a longitudinal vector, and
we have made use of the fact that the linearized energy-momentum
tensor of the fields $B^\mu$ and $\phi$ has zero anisotropic
stress.

The complete expressions for the quadratic action and linear
equations for
perturbations in the slow roll regime (i.e., neglecting terms suppressed
by $X'',~X',~\varphi'',~\varphi'$;
hereafter prime denotes $\frac{\d}{\d \eta}$) are given in
Appendix for completeness. However, we are interested in momenta
$P \sim \epsilon$.
According to the
relations (\ref{approx}),
modes of interest have
relatively high spatial momenta and frequencies,
\be
   P,\; \Omega \gg   m, \; M \; , \;\;\;\;\;  P,\; \Omega \gg H
\label{mar15-2}
\ee
where $\Omega $  is the physical
frequency.
Note that in the
regime (\ref{mar15-1})
one has $H \sim M$, so the second of these
inequalities is a consequence of the first one. For momenta obeying
(\ref{mar15-2}), the equations simplify considerably:
\bea
\chi''-\Delta\chi + \epsilon ab'_0 -\epsilon a\partial_ib_i-2\epsilon
a^2X\Phi' &=& 0
\label{eq-chi-m} \\
\epsilon a\chi' -\gamma\left(b''_0-\Delta b_0\right)
 +\gamma aX(\Phi''-\Delta\Phi)
&=& 0
\label{eq-b0-m}\\
\epsilon a\partial_i\chi+\alpha\left(b''_i-\Delta
b_i \right)
-2\alpha aX\partial_i\Phi'&=& 0
\label{eq-bi-m}\\
-\frac{3M_{PL}^2}{4\pi} a (\Phi''-\Delta\Phi)
-2\epsilon
aX \chi' + \gamma X(b''_0-\Delta b_0)
+2\alpha X\partial_ib'_i
&=& 0
\label{eq-Phi-m}
\eea
We are now in a position to solve these
equations in the WKB approximation by writing
\[
   \{\chi, b_0, b_i, \Phi \} \propto
  \mbox{exp} \left( i \int \omega d\eta - i {\bf p x} \right)
\]
with slowly varying amplitudes. Before doing that, we note that
at $P, \Omega \sim \epsilon$,
one estimates from  eq.~(\ref{eq-Phi-m})
\[
\Phi \sim \frac{X}{M_{Pl}^2 a} \{ \chi, b_0, b_i \} \; .
\]
Therefore, the last terms in eqs.~(\ref{eq-chi-m}),
(\ref{eq-b0-m}) and (\ref{eq-bi-m}), describing back reaction of
the gravitational potential on the field perturbations, are
suppressed as compared to other terms by  $X^2
/ M_{Pl}^2 \sim m^2/\epsilon^2$. Neglecting these terms, we arrive
at the system of equations for dark  energy
perturbations, which we write in the leading order of the WKB
approximation, \bea (\omega^2 - p^2)\chi- i\epsilon a\omega
b_0+i\epsilon apb_L &=& 0 \; ,
\nonumber \\
i\epsilon a\omega\chi+\gamma (\omega^2 - p^2)
b_0 &=& 0 \; ,
\nonumber\\
i\epsilon ap\chi-\alpha (\omega^2 -p^2) b_L &=&0 \; ,
\nonumber
\eea
where $b_L = (p_i/p) b_i$.
This is basically the same system as that obtained in Minkowski
space-time. There are three modes, two of which are normal at all
momenta, while the third  one is tachyonic at low momenta.
In terms of physical frequency, the dispersion
relation for the latter
mode reads
\be
\Omega^2
\equiv \frac{\omega^2}{a^2}
=P^2+\frac{\epsilon^2}{2\gamma}\left(1-
\sqrt{1+\frac{4\zeta \gamma^2}{\epsilon^2} P^2}\right) \; ,
\label{mar17-6}
\ee
where
\[
\zeta = \frac{1}{\alpha} + \frac{1}{\gamma} \; .
\]
As promised, the tachyonic regime occurs at
\[
P < P_c \equiv \frac{\epsilon}{\sqrt{\alpha}} \; .
\]
The corresponding solution at $P>P_c$ is given by
\bea
\chi &=& C\cdot \frac{i (\omega^2 - p^2)}{\epsilon \omega}\, f(\eta) \,
\mbox{e}^{i \frac{\pi}{4}}\; ,
\nonumber \\
b_0 &=& C\cdot \frac{a}{\gamma} \, f(\eta) \, \mbox{e}^{i \frac{\pi}{4}}\; ,
\nonumber \\
b_L &=& - C \cdot
\frac{ap}{\alpha \omega} \, f(\eta) \, \mbox{e}^{i \frac{\pi}{4}}\; ,
\label{mar17-1}
\eea
where $f(\eta)$ is a slowly varying function, $C$ is an overall constant
 and
 the phase factor is introduced for convenience.

The function $f(\eta)$ can be found by noting that
with our approximations, the action for the field perturbations
reads
\be
S=\int d\eta
d \textbf{x}\left[\frac{\alpha}{2}((b'_i)^2-(\d_i b_j)^2)
+\frac{\gamma}{2}((b'_0)^2-({\bf \nabla}
b_0)^2)+\frac{1}{2}((\chi')^2-({\bf \nabla}\chi)^2)+\epsilon
a(\chi'b_0-\partial_i\chi b_i)\right] \; .
\label{mar17-3}
\ee
The corresponding energy functional is
\[
E=\int d\textbf{x} \left[\frac{\alpha}{2}((b'_i)^2+
(\d_i b_j)^2)+\frac{\gamma}{2}((b'_0)^2+({\bf \nabla}
b_0)^2)+\frac{1}{2}((\chi')^2+({\bf \nabla}\chi)^2)+\epsilon
a\partial_i\chi b_i\right] \; .
\]
In the WKB approximation, this energy is conserved. Making use of the
expressions (\ref{mar17-1}) and requiring that energy of this solution
is conserved,
one finds 
\be f(\eta) = \left\vert \frac{\omega}{2\left(\frac{
(\omega^2-p^2)^2}{\epsilon^2}+ \zeta
a^2p^2\right)}\right\vert^{\frac{1}{2}} \; , \label{mar17-2} \ee
The same result can be obtained by considering the conservation of
the Wronskian of the  system (\ref{eq-chi-m}),
(\ref{eq-b0-m}), (\ref{eq-bi-m}), which, with our approximations,
is
\[
W = -i \left(\chi^* \chi^\prime + \gamma b_0^* b_0^\prime
+ \alpha b_L^* b_L^\prime + \epsilon a \chi^* b_0 - \mbox{c.c.} \right) \; .
\]

Now, to find the overall constant in (\ref{mar17-1}) we quantize the
system with the action (\ref{mar17-3}) at early times and obtain
\be
C = \frac{A^+_{\bf p}}{(2\pi)^{3/2}} \; ,
\label{mar17-4}
\ee
where the creation and annihilation operators obey the standard
commutational relation $[A^-_{\bf p}, A^+_{\bf p'}] = \delta ({\bf p} -
{\bf p'})$.

We are interested in the behavior of perturbations in the
tachyonic regime. The exponentially growing part is found by the
standard WKB continuation of the expressions (\ref{mar17-1}),
(\ref{mar17-2}) and (\ref{mar17-4}) ``beyond the turning point''.
In this way we finally obtain in the tachyonic 
regime \bea \chi &=& \int \frac{d\textbf{p}}{(2\pi)^{3/2}}
\left(\frac{i(\omega^2-p^2)}{\epsilon\omega}\, f \,
A^+_{\textbf{p}} \mbox{e}^{-i{\bf px}} +\mbox{h.c.} \right)
\mbox{e}^{\int |\omega| d\eta} \; ,
\nonumber \\
b_0 &=& \int \frac{d\textbf{p}}{(2\pi)^{3/2}}
\left(\frac{a}{\gamma}\, f \, A^+_{\textbf{p}}\mbox{e}^{-i{\bf
px}}+ \mbox{h.c.} \right) \mbox{e}^{\int |\omega| d\eta} \; ,
\nonumber \\
b_i &=& \int \frac{d\textbf{p}}{(2\pi)^{3/2}} \left(-\frac{p_ia
}{\alpha\omega} \, f \, A^+_{\textbf{p}} \mbox{e}^{-i{\bf px}}+
\mbox{h.c.} \right) \mbox{e}^{\int |\omega| d\eta} \; , \nonumber
\eea where $f(\eta)$ is still given by (\ref{mar17-2}).

\section{Gravitational potentials}

The gravitational potential $\Phi$ is
determined by eq.~(\ref{eq-Phi-m}). After straightforward
calculation we  obtain that at $P>P_c$ the
explicit expression is
\[
\Phi=- \sqrt{\frac{4\pi}{27}}\frac{m}{M_{PL}\epsilon}
\left(\frac{4\gamma P^2}{\epsilon^2}
\frac{1}{\sqrt{1+ \frac{4\zeta \gamma^2}{\epsilon^2} P^2} -1} -1
\right) \cdot C \cdot f(\eta)\, \mbox{e}^{i\frac{\pi}{4}} \; ,
\]
where $C$ and $f(\eta)$ are the same as in
(\ref{mar17-1}). Continuing into the tachyonic region, we obtain
\[
\Phi = - \sqrt{\frac{4\pi}{27}}\frac{m}{M_{PL}\epsilon} \int
\frac{d\textbf{p}}{(2\pi)^{3/2}} \left[ \left(\frac{4\gamma
P^2}{\epsilon^2} \frac{1}{\sqrt{1+ \frac{4\zeta
\gamma^2}{\epsilon^2} P^2} -1} -1 \right) \, f \, A^+_{\textbf{p}}
\mbox{e}^{-i{\bf px}}+ \mbox{h.c.} \right] \mbox{e}^{\int |\omega|
d\eta} \; .
\]
This expression gets simplified towards the end of the development
of instability, i.e. at $P \ll \epsilon$ (but still $P \gg H\sim M$).
In that case the dispersion relation (\ref{mar17-6}) reads
\[
   \omega^2 = - \frac{\gamma}{\alpha} p^2 \; ,
\]
so that
\[
   f (\eta) = \frac{\gamma^{3/4} \alpha^{1/4}}{(\alpha + \gamma)^{1/2}}
\frac{1}{\sqrt{2p} a(\eta)} \; .
\]
Hence
\[
\Phi = - \sqrt{\frac{4\pi}{27}}\frac{m}{M_{PL}\epsilon}
\frac{\gamma^{3/4} \alpha^{1/4} (\alpha - \gamma)}{(\alpha + \gamma)^{3/2}}
\int
\frac{d\textbf{p}}{(2\pi)^{3/2}} \frac{1}{\sqrt{2p} a}
 \left( A^+_{\textbf{p}} \mbox{e}^{-i{\bf px}}+ \mbox{h.c.}
\right)  \mbox{e}^{\int |\omega| d\eta} \; .
\]
This is Gaussian random field whose variance we write in terms of
the physical momentum and frequency:
\be
\langle \Phi^2 ({\bf x}, t) \rangle
= \frac{1}{27 \pi}
\frac{\gamma^{3/2} \alpha^{1/2} (\alpha - \gamma)^2}{(\alpha + \gamma)^{3}}
\left( \frac{m}{M_{PL}\epsilon} \right)^2
\int P dP~   \mbox{exp} \left(2\int_{t_c}^{t} |\Omega| d t \right) \; ,
\label{mar17-8}
\ee
where the moment $t_c$ corresponds to the beginning of the
tachyonic regime, i.e.,
\[
   P(t_c) \equiv \frac{p}{a(t_c)} =P_c
\; .
\]
Note that the exponential growth factor is large but finite in the formal
limit $t \to \infty$. As an example, for constant $H$, the
total growth factor is
\[
\int_{t_c}^\infty \Omega(t)dt
=\frac{1}{H}\int_0^{P_c}\Omega(P)\frac{dP}{P}=
d(\alpha,\gamma)\cdot\frac{\epsilon}{H} \; ,
\]
where
$d(\alpha,\gamma)$ is of
order one (for example, $d(1,\frac{1}{2})\approx 0.499$).

\section{Discussion}

Recalling the relations (\ref{approx}) and (\ref{mar15-1})
one observes that the gravitational potentials grow from very small
values, set by vacuum fluctuations at the low momentum scale
$P_c \sim \epsilon$, but they eventually may become large.
On the one hand, as discussed in Ref.~\cite{Libanov:2007mq}, this
gives rise  to a constraint on the parameters of the model we discuss
in this paper: by requiring that $\Phi \ll 1$ at the present
time, one obtains
\[
\epsilon\lesssim\frac{1}{d(\alpha,\gamma)} H_0 \ln\left(\frac{
M_{PL}}{H_0}\right)\simeq 280 H_0 \;\;\; \mbox{for} \;\;\;
\alpha = 1\; , \;\;\; \gamma = 0.5 \; ,
\]
where $H_0$ is the present value of the Hubble parameter. So, the
model may serve as a viable  description of
dark energy at the expense of extra fine tuning, over and beyond
the usual fine tuning required to get the right value for cosmic
acceleration. On the other hand, with even more fine tuning, the
dark energy induced gravitational potentials may be sizeable (and
still acceptably small) today. A novel feature here is that the
emergence of these potentials would be a fairly recent phenomenon:
the largest values of the exponential factor in (\ref{mar17-8})
are obtained at late times when $\Omega(t)$ changes slowly. Also,
the amplitude of perturbations would be peaked at a certain
momentum $P$: for high momenta, the tachyonic regime has not yet
started, while for low momenta the integral in the exponent is
saturated at early times, and hence is small. These properties may
be potentially observable.

The authors are indebted to O.~Khovanskaya, M.~Libanov 
and M.~Sazhin for helpful discussions. This work
is supported in part by Russian Foundation for Basic
Research, grant 08-02-00473.

\section*{Appendix. Complete action and equations for perturbations}

For completeness, we present here the complete
expressions for the quadratic action and linear equations for the
scalar
perturbations, valid in the slow roll regime. We use the conformal
Newtonian gauge and neglect the terms suppressed by
$X'',~X',~\varphi'',~\varphi'$. The quadratic action is
\begin{equation}
S^{(2)}=\int d\eta d\textbf{x}~a^4L^{(2)},
\label{mar17-9}
\end{equation}
where
\begin{equation}
\begin{aligned}
a^4L^{(2)}=-\frac{M_{PL}^2}{\pi}\left(- \frac{3}{8}a^2\Phi\Delta\Phi-\frac{3}{2}a^3H\Phi\Phi'-\frac{3}{2}aa''\Phi^2-\frac{3}{8}a^2(\Phi')^2 \right)-
\\ -\frac{\alpha}{2}\left[-\left(b'_i-aX\partial_i\Phi-aHb_i\right)^2-a^2H^2b_i^2+2aH(\partial_ib_0-aX\partial_i\Phi)b_i\right]-
\\ -\frac{\alpha}{2}\left[(\partial_ib_j)^2-2a\partial_ib_i(Hb_0-X\Phi')+3(-aHb_0+4a'X\Phi+aX\Phi')^2\right]-
\\ -\alpha a'X\left(\partial_i\Phi
b_i-3b_0\Phi'-6a'X\Phi^2-6aX\Phi\Phi'\right)+
\\ +\frac{\gamma}{2}\left[\left(b'_0-aHb_0-aX\Phi'\right)^2-\left(\partial_i
b_0-aX\partial_i\Phi\right)^2\right]+
\\ +\frac{1}{2}\left(\chi'-\chi
aH\right)^2-\frac{1}{2}\left(\partial_i\chi\right)^2+\epsilon
a\left(\chi'-\chi aH\right)\left(b_0-2aX\Phi\right)-\epsilon
a\partial_i\chi b_i+
\\
+\frac{M^2a^2}{2}\left(b_0^2-4aXb_0\Phi+4a^2X^2\Phi^2-b_i^2\right)-\frac{m^2a^2}{2}\chi^2.
\end{aligned}
\nonumber
\end{equation}
From here we obtain the equations for the field perturbations:
\begin{equation} \label{pert}
\begin{aligned}
\chi''-\Delta\chi-\frac{a''}{a}\chi+m^2a^2\chi+\epsilon
ab'_0+2\epsilon a'b_0-\epsilon a\partial_ib_i-2\epsilon
a^2X\Phi'-6\epsilon aa'X\Phi=0
\\
\\
\epsilon(a\chi'-a'\chi)-\gamma\left(b''_0-\Delta b_0-\frac{a''}{a}b_0\right)+(M^2-3\alpha H^2)a^2b_0+2\alpha
aH\partial_ib_i+
\\ +\gamma aX(\Phi''-\Delta\Phi)+2(3\alpha+\gamma)a'X\Phi'+2(6\alpha H^2-M^2)a^3X\Phi=0
\\
\\
\epsilon a\partial_i\chi+2\alpha
aH\partial_ib_0+\alpha\left(b''_i-\Delta
b_i-\frac{a''}{a}b_i\right)+\left(M^2-\alpha
H^2\right)a^2b_i-2\alpha aX\partial_i\Phi'-2\alpha
a^2HX\partial_i\Phi=0
\\
\\
-\frac{3M_{PL}^2}{4\pi}\left[a^2(\Phi''-\Delta\Phi)+2aa'\Phi'-2(aa''-(a')^2)\Phi\right]-2\epsilon
aX(a\chi'-a'\chi)+\gamma aX(b''_0-\Delta b_0)-
\\ -6\alpha a^2HXb'_0-\gamma a''Xb_0-2M^2a^3Xb_0+12\alpha a^3H^2Xb_0-6\alpha
a''Xb_0+2\alpha aX\partial_ib'_i+
\\
+a^2X^2(3\alpha-\gamma)\Phi''+(\gamma-\alpha)a^2X^2\Delta\Phi+2(3\alpha-\gamma)a^3HX^2\Phi'+6\alpha(aa''-5(a')^2)X^2\Phi+
\\ +4M^2a^4X^2\Phi=0
\end{aligned}
\nonumber
\end{equation}
Equations (\ref{eq-chi-m}), (\ref{eq-b0-m}), (\ref{eq-bi-m})
and (\ref{eq-Phi-m}) follow from the latter equations in the regime
(\ref{mar15-2}), while the action (\ref{mar17-9})
reduces to (\ref{mar17-3}) in this regime.


\end{document}